\documentclass[journal, 12pt, draftcls, onecolumn]{IEEEtran}
\usepackage{cite}
\usepackage{footnote}
\usepackage{xcolor}
\usepackage{amsmath,amsfonts,amssymb,graphicx,eqnarray}
\usepackage{graphicx}
\usepackage{ifpdf}
\usepackage{amsmath}
\usepackage{amsthm}
\newtheorem{theorem}{Theorem}
\newtheorem{corollary}{Corollary}
\newtheorem{lemma}{Lemma}

\ifCLASSINFOpdf
\else
\fi

\begin{document}
\bibliographystyle{IEEEtran}
\title{Massive MIMO for Wireless Sensing \\ with a Coherent Multiple Access Channel}

\author{\IEEEauthorblockN{Feng Jiang$^{1}$, Jie Chen$^{1}$, A. Lee Swindlehurst$^{1}$ and Jos\'{e} A. L\'{o}pez-Salcedo$^{2}$\\ \IEEEauthorblockA{$^1$Department of Electrical Engineering and Computer Science\\University of California at Irvine, Irvine, CA 92697, USA\\$^2$Department of Telecommunications and Systems Engineering\\ Universitat Aut\`{o}noma de Barcelona (UAB), 08193 Bellaterra (Barcelona), Spain\\Email:\{feng.jiang, jie.chen, swindle\}@uci.edu, jose.salcedo@uab.es}}}

%


\maketitle

\begin{abstract}
We consider the detection and estimation of a zero-mean Gaussian signal in a wireless sensor network with a coherent multiple access channel, when the fusion center (FC) is configured with a large number of antennas and the wireless channels between the sensor nodes and FC experience Rayleigh fading. For the detection problem, we study the Neyman-Pearson (NP) Detector and Energy Detector (ED), and find optimal values for the sensor transmission gains.  For the NP detector which requires channel state information (CSI), we show that detection performance remains asymptotically constant with the number of FC antennas if the sensor transmit power decreases proportionally with the increase in the number of antennas. Performance bounds show that the benefit of multiple antennas at the FC disappears as the transmit power grows.  The results of the NP detector are also generalized to the linear minimum mean squared error estimator. For the ED which does not require CSI, we derive optimal gains that maximize the deflection coefficient of the detector, and we show that a constant deflection can be asymptotically achieved if the sensor transmit power scales as the inverse square root of the number of FC antennas.  Unlike the NP detector, for high sensor power the multi-antenna ED is observed to empirically have significantly better performance than the single-antenna implementation. A number of simulation results are included to validate the analysis.

\end{abstract}


%
\setlength{\baselineskip}{0.98\baselineskip}
\section{Introduction}
\subsection{Background}
The use of wireless sensor networks (WSNs) for detection and parameter estimation has been widely studied ({\em e.g.},\cite{Viswanathan:1997,  Niu:2006, Chamberland:2003, Chamberland:2004,  Cui:2007, Li:2007, Gastpar:2008, Quan:2010, Leong:2011, Jiang:2013, Zhang:2013}) . When a coherent multiple access channel is employed between the sensor nodes and fusion center (FC) \cite{Chamberland:2003, Chamberland:2004, Cui:2007, Li:2007,Gastpar:2008, Quan:2010, Leong:2011, Jiang:2013}, each sensor takes a noisy measurement of the signal of interest, amplifies and forwards the measurement to a FC through a wireless fading channel, and the FC makes a decision about the presence of the signal and estimates its parameters based on the coherent sum of the signals from all the sensor nodes. To minimize the detection or estimation errors, the transmit power at the sensors is optimized under either sum or individual power constraints.
The aforementioned works all assume that the FC is configured with a single antenna. It is well-known that multiple antennas can effectively increase the throughput of a wireless link, and recently researchers have investigated the use of arrays with a massive number of antennas in wireless communication systems in order to improve spectral and energy efficiency \cite{Marzetta:2010, Ngo:2013, Hoydis:2013, Lu:2014}.  Most of the research on so-called ``massive MIMO'' systems has been focused on cellular networks where the base station (BS) is configured with many antennas while the individual mobile stations have a single antenna.  When perfect channel state information (CSI) is available at the BS, it has been shown that the transmit power of the mobile terminals can be reduced proportionally to the increase in the number of antennas without impacting the asymptotic rate of the users in the system \cite{Marzetta:2010}. The benefit is somewhat less when the BS uses an imperfect channel estimate; in this case the mobile users' transmit power can be inversely proportional to the square root of the number of antennas in order to achieve a constant rate \cite{Ngo:2013}.

For parameter detection or estimation problems in WSNs, an important question is how to exploit a multi-antenna FC to improve the probability of detection or estimation error. Several recent papers have studied the benefit provided by multiple antennas in the WSN context \cite{Nevat:2014, Smith:2009, Banavar:2012, Ciuonzo:2012, Jiang:20122, Ciuonzo:2015}.  In \cite{Nevat:2014}, the sensors use a fixed transmission gain to forward the measured signal to the multi-antenna FC, and the probabilities of detection and false alarm are derived under different assumptions for the CSI.  Power allocation problems for signal detection and estimation are formulated in \cite{Smith:2009, Banavar:2012} for a multi-antenna FC under a Rayleigh fading channel, but the performance benefit of a multiple- versus single-antenna FC is shown to be bounded by a constant that is unrelated to the number of antennas.  For signal estimation using a phase-shift and forward WSN with a multi-antenna FC, it has been shown in \cite{Jiang:20122} that as the number of antennas $M$ grows large, in certain cases the estimation error will decrease by a factor of $M$. Antenna arrays at the FC are also considered in \cite{Ciuonzo:2012, Ciuonzo:2015}, where each sensor node first makes a local binary decision about the measured signal, and then forwards the decisions to the multi-antenna FC using uniform transmit power. In \cite {Ciuonzo:2012}, a number of sub-optimal but low complexity fusion rules at the FC are derived and analyzed, and the results indicate the benefit of using multiple antennas in terms of detection performance. The recent work in \cite{Ciuonzo:2015} shows that when the number of FC antennas is very large, low complexity algorithms can asymptotically achieve an upper bound on detection performance even using a linear receiver with imperfect CSI.

While the benefits of massive numbers of antennas have been carefully studied for communication systems, we see above that relatively little work has analyzed their impact for WSNs.  In this paper, we investigate the gains in energy efficiency that can be obtained in a coherent multiple-access WSN when the FC has a large number of antennas, and we show how to determine optimal values for the sensor gains when the CSI is either perfectly known or unknown at the FC.  In particular, our motivation is to demonstrate that FC antennas can be traded for sensor power; this is an important observation for WSNs where the sensors must conserve energy (\emph{e.g.}, due to the use of batteries or energy harvesting). The specific contributions of the paper are detailed in the next section.

\subsection{Contributions}
In this paper, we study the detection and estimation performance of a coherent amplify-and-forward WSN with single antenna sensors and a massive number $M$ of antennas at the FC. We assume the parameter of interest is a zero-mean circular complex Gaussian variable and that the wireless channels between the sensor nodes and FC undergo Rayleigh fading. Under these assumptions, we investigate the performance of the Neyman-Pearson (NP) and energy detectors (ED) and the linear minimum mean squared error estimator (LMMSE).  Our contributions are summarized below.

(1) For the case where CSI for the sensor nodes is available at the FC and the NP detector can be implemented, we derive the dependence of both probability of detection (PD) and probability of false alarm (PFA) on the sensor transmit power and show that as $M\rightarrow\infty$, the sensor power can be reduced by $1/M$ to achieve a constant PD for the same fixed PFA. This is similar in spirit to the results for massive MIMO in wireless cellular communications with perfect CSI \cite{Ngo:2013}. However, unlike \cite{Ngo:2013} which assumes each user transmits with equal power, we derive the optimal transmission gains for the sensors that maximize PD for a fixed PFA under a sum power constraint.  We show that this problem is independent of the sensor phase and convex with respect to the magnitude squared of the sensor gain as $M \rightarrow \infty$, and we formulate a simple closed-form ``water-filling'' solution to calculate the optimal gains. In our simulations, we demonstrate that compared with a uniform power allocation, the optimal gains result in significantly improved PD performance when the sensors transmit with low power, which is the case of interest for energy efficiency.

(2) For the NP detector, we also derive asymptotic performance bounds for cases where the available sum transmit power $P$ satisfies either $P \rightarrow\infty$ or $P\rightarrow 0$.  When $P \rightarrow 0$, we show that PD approaches PFA in the single antenna case, but PD is strictly greater than PFA (and potentially significantly greater than PFA) as long as $P$ decreases at a rate of $O(1/M)$ or slower as $M \rightarrow\infty$.  However, when $P\rightarrow\infty$, we show that both the single- and multiple-antenna FCs asymptotically achieve the same detection performance, and hence the use of multiple antennas asymptotically provides no benefit for the NP detector at very high signal-to-noise ratios.

(3) For the case where the CSI is unknown or a computationally simpler solution is desired, we study the performance of the ED. The \emph{deflection} of the ED is used as the performance metric, which generally serves as an accurate indicator of a detector's performance. Our results show that if the sensor transmit power decreases as $1/\sqrt{M}$ when $M \rightarrow\infty$, a constant deflection can be achieved.  Based on this, we show how to choose the sensor transmission gains to maximize the deflection under a sum power constraint.  In particular, we show that when $M \rightarrow\infty$, the optimal gains can be found in the general case via a quadratically constrained linear program, and we also show that closed-form solutions are possible for limiting values of the power constraint $P$. As in the NP detector case, the optimal solution is independent of the sensor phase. Simulation results demonstrate that reducing transmit power by $1/\sqrt{M}$ to maintain a constant deflection as $M$ grows results in a constant PD. Note that although this result is superficially similar to a result in \cite{Ngo:2013}, the case we consider is considerably different since it involves the energy detector which requires no CSI, unlike \cite{Ngo:2013} which assumes a minimum-mean squared error channel estimate obtained using pilot signals.  Also, unlike \cite{Ngo:2013}, we do not assume a uniform power allocation, but as mentioned above we instead derive optimal sensor transmit gains and illustrate when these optimal gains provide significantly better detection performance.

(4) For the LMMSE estimator, we prove that a constant MSE can be achieved by decreasing the transmit power as $1/M$ as the number of FC antennas $M$ grows. This result is obtained by generalizing the asymptotic results for the NP detector to the LMMSE estimator, and showing that the PD of the NP detector and the LMMSE mean-squared error (MSE) both obey a similar rule as $M \rightarrow \infty$. We also derive bounds on the MSE for the limiting cases $P \rightarrow 0$ and $P \rightarrow \infty$, and show similar behavior for these bounds as in the case of PD for the NP detector.

Some of the contributions listed above appeared previously in the conference paper \cite{Jiang:2014}.

\subsection{Organization}
The remainder of the paper is organized as follows. In Section \ref{sec:two}, we introduce the signal model and derive basic results for PD and PFA.  In Section \ref{sec:three}, we prove the main results for the NP detector and LMMSE estimator, and we formulate and solve the sensor transmission gain optimization problem to maximize PD for a given PFA under a sum transmit power constraint. The deflection of the energy detector is analyzed in Section~{\ref{sec:four}}, and the problem of calculating the transmission gains that maximize the deflection is solved. The results of several simulation studies are provided in Section~\ref{sec:six} to validate the theoretical derivations, and the conclusions of the paper are summarized in Section~\ref{sec:seven}.

The notation used in this paper is summarized as follows. Lower-case and upper-case bold letters represent vectors and matrices respectively, and $\mathbb{C}^{M\times 1}$ denotes the space of $M$-element complex vectors. We use $(\cdot)^T$ and $(\cdot)^H$ for transpose and conjugate transpose respectively. The $M\times M$ identity matrix is denoted as $\mathbf{I}_M$ and $\mathrm{diag}\{d_1 \; \cdots \; d_N\}$ is a $N\times N$ diagonal matrix with $d_i$ as the $i$th diagonal element. Probabilities and conditional probabilities are denoted by $\mathrm{Pr}(\cdot)$ and $\mathrm{Pr}(\cdot|\cdot)$, and $p(\cdot|\cdot)$ represents a conditional probability density function. The functions $\mathbb{E}\{\cdot\}$ and $\mathrm{Var}\{\cdot\}$ denote the expectation and variance of a random variable, and $\mathcal{CN}(\mathbf{0},\mathbf{\Sigma})$ denotes the complex Gaussian distribution with zero mean and covariance matrix $\mathbf{\Sigma}$.  The $i$th eigenvalue of a matrix is written as $\lambda_i(\cdot)$, and for two Hermitian matrices $\mathbf{A}$ and $\mathbf{B}$, $\mathbf{A}\succeq\mathbf{B}$ means that $\mathbf{A}-\mathbf{B}$ is positive semidefinite. 

\section{Signal Model and Neyman Pearson Detector}\label{sec:two}
We consider a general binary Gaussian detection problem, where the signal of interest $\theta$ is modeled as a zero-mean circular complex Gaussian variable\footnote{ Although we use a Bayesian framework, our approach can be also used for the deterministic case, in which $\theta$ is assumed to be a deterministic signal.} with variance $\sigma_{\theta}^2$, a distribution we denote by $\mathcal{CN}(0,\sigma_{\theta}^2)$. The measurement available at the $i$th of $N$ sensor nodes is given by
\begin{equation}
s_i=\theta+v_i\;,
\end{equation}
where $v_i$ is measurement noise distributed as $\mathcal{CN}(0,\sigma_{v,i}^2)$. The $i$th sensor multiplies the measurement with a complex gain $a_i$ and coherently forwards the result to the FC through a wireless fading channel. The received signal at the $M$-antenna FC under the two hypotheses is
\begin{subequations}
\begin{align}
\mathcal{H}_0: \mathbf{y}&=\mathbf{H}\mathbf{D}\mathbf{v}+\mathbf{n}\label{eq:rsig1}\\
\mathcal{H}_1: \mathbf{y}&=\mathbf{H}\mathbf{a}\theta+\mathbf{H}\mathbf{D}\mathbf{v}+\mathbf{n}\label{eq:rsig2}\;,
\end{align}
\end{subequations}
where
\begin{subequations}
\begin{align}
\mathbf{v} &= [v_{1} \; \cdots \; v_N]^T  \IEEEyessubnumber\\
\mathbf{a} &= [a_1 \; \cdots \; a_N]^T  \\
\mathbf{D} &= \mathrm{diag}\{a_1 \; \cdots \; a_N\} \\
\mathbf{H} &= [\mathbf{h}_1 \; \cdots \; \mathbf{h}_N] \; ,
\end{align}
\end{subequations}$\mathbf{h}_i\in\mathbb{C}^{M\times 1}$ is the channel gain between the $i$th sensor and the FC, and the vector $\mathbf{n}\in\mathbb{C}^{M\times 1}$ represents additive Gaussian noise at the FC and has the distribution $\mathcal{CN}(\mathbf{0},\sigma^2_n\mathbf{I}_{M})$.

Assuming that the FC has perfect knowledge of signal variance $\sigma_{\theta}^2$, the measurement noise power $\sigma_{v,i}^2$ and the CSI in ${\mathbf{H}}$, the NP criterion can be used to distinguish between the hypotheses $\mathcal{H}_0$ and $\mathcal{H}_1$. The NP detector decides $\mathcal{H}_1$ if \cite{Kay:19932}
\begin{equation}\label{eq:lr}
L(\mathbf{y})=\frac{p(\mathbf{y}|\mathcal{H}_1)}{p(\mathbf{y}|\mathcal{H}_0)}>\gamma
\end{equation}
for a given threshold $\gamma$, where $p(\mathbf{y}|\mathcal{H}_1)$ and $p(\mathbf{y}|\mathcal{H}_0)$ are the conditional probability density functions (PDFs) of $\mathbf{y}$ under $\mathcal{H}_1$ and  $\mathcal{H}_0$, respectively. Assume the measurement noise at the sensors is independent, so that the covariance of $\mathbf{v}$ is given by $\mathbf{V}=\mathrm{diag}\{\sigma_{v,1}^2 \; \cdots \; \sigma_{v,N}^2\}$. Since $\mathbf{y}$ is Gaussian under both $\mathcal{H}_1$ and $\mathcal{H}_0$, we have \cite{Kay:19932}
\begin{subequations}
\begin{align}
p(\mathbf{y}|\mathcal{H}_1)&=\frac{1}{\pi^M\mathrm{det}(\mathbf{C}_s\!+\!\mathbf{C}_w)}\mathrm{exp}\left(-\mathbf{y}^H(\mathbf{C}_s\!+\!\mathbf{C}_w)^{-1}\mathbf{y}\right)\label{eq:pdf1}\\
p(\mathbf{y}|\mathcal{H}_0)&=\frac{1}{\pi^M\mathrm{det}(\mathbf{C}_w)}\mathrm{exp}\left(-\mathbf{y}^H\mathbf{C}_w^{-1}\mathbf{y}\right)\;\label{eq:pdf2},
\end{align}
\end{subequations}
where $\mathbf{C}_w=\mathbf{H}\mathbf{D}\mathbf{V}\mathbf{D}^{H}\mathbf{H}^{H}+\sigma^2_n\mathbf{I}_M$ is the covariance of $\mathbf{y}$ under $\mathcal{H}_0$, $\mathbf{C}_s=\sigma_{\theta}^2\mathbf{H}\mathbf{a}\mathbf{a}^H\mathbf{H}^H$ and $\mathbf{C}_w+\mathbf{C}_s$ is the covariance of $\mathbf{y}$ under $\mathcal{H}_1$.

\begin{lemma}
Based on the signal model in (\ref{eq:rsig1}) and (\ref{eq:rsig2}), and the conditional PDFs in (\ref{eq:pdf1}) and (\ref{eq:pdf2}), the NP detector in (\ref{eq:lr}) is equivalent to deciding $\mathcal{H}_1$ if
\begin{equation}\label{eq:lemma1}
\sigma_{\theta}^2|\mathbf{a}^H\mathbf{H}^H\mathbf{C}_w^{-1}\mathbf{y}|^2>\gamma^{'}\;,
\end{equation}
where
\begin{align}
\gamma^{'}&=(1+\sigma_{\theta}^2g(\mathbf{a})) \ln\left[\gamma(1+\sigma_{\theta}^2g(\mathbf{a}))\right] \\
g(\mathbf{a})&=\mathbf{a}^H\mathbf{H}^H\mathbf{C}_w^{-1}\mathbf{H}\mathbf{a} \; .
\end{align}
\end{lemma}
\begin{IEEEproof}
See Appendix \ref{app:lemma1}\;.
\end{IEEEproof}

For the NP detector in (\ref{eq:lemma1}), the probability of detection $P_{D}$ and probability of false alarm $P_{FA}$ are defined as
\begin{subequations}
\begin{align}
P_{D}&=\mathrm{Pr}\left(\sigma_{\theta}^2|\mathbf{a}^H\mathbf{H}^H\mathbf{C}_w^{-1}\mathbf{y}|^2>\gamma^{'}|\mathcal{H}_1\right)\\
P_{FA}&=\mathrm{Pr}\left(\sigma_{\theta}^2|\mathbf{a}^H\mathbf{H}^H\mathbf{C}_w^{-1}\mathbf{y}|^2>\gamma^{'}|\mathcal{H}_0\right)\;.
\end{align}
\end{subequations}
To evaluate $P_{D}$, we first rewrite it as
\begin{equation}\label{eq:pd}
P_{D}=\mathrm{Pr}\left(\sigma_{\theta}^2\tilde{\mathbf{y}}^H\mathbf{W}\tilde{\mathbf{y}}>\gamma^{'}|\mathcal{H}_1\right)\;,
\end{equation}
where $\tilde{\mathbf{y}}=(\mathbf{C}_s+\mathbf{C}_w)^{-\frac{1}{2}}\mathbf{y}$ and
\begin{equation}
\mathbf{W}=(\mathbf{C}_s+\mathbf{C}_w)^{\frac{1}{2}}\mathbf{C}_w^{-1}\mathbf{H}\mathbf{a}\mathbf{a}^H\mathbf{H}^H\mathbf{C}_w^{-1}(\mathbf{C}_s+\mathbf{C}_w)^{\frac{1}{2}}\;\nonumber.
\end{equation}
Since $\mathbf{y}\sim\mathcal{CN}(\mathbf{0},\mathbf{C}_s+\mathbf{C}_w)$ under $\mathcal{H}_1$,  $\tilde{\mathbf{y}}=(\mathbf{C}_s+\mathbf{C}_w)^{-\frac{1}{2}}{\mathbf{y}}$ is distributed as  $\mathcal{CN}(\mathbf{0},\mathbf{I}_M)$. Defining the eigendecomposition of $\mathbf{W}$ as
\begin{equation}
\mathbf{W}=\mathbf{U}\mathbf{G}\mathbf{U}^H\;
\end{equation}where $\mathbf{G}=\mathrm{diag}\{g(\mathbf{a})+\sigma_{\theta}^2g(\mathbf{a})^2,0\cdots 0\}$\;, equation~(\ref{eq:pd}) becomes
\begin{align}\label{eq:pd2}
P_{D}&=\mathrm{Pr}\left(\sigma_{\theta}^2\tilde{\mathbf{y}}^H\mathbf{U}\mathbf{G}\mathbf{U}^H\tilde{\mathbf{y}}>\gamma^{'}|\mathcal{H}_1\right)\nonumber\\
&\overset{(b)}{=}\mathrm{Pr}\left(\sigma_{\theta}^2\tilde{\mathbf{y}}^H\mathbf{G}\tilde{\mathbf{y}}>\gamma^{'}|\mathcal{H}_1\right)\nonumber\\
&\overset{(c)}{=}\exp\left(-\frac{\gamma^{'}}{{\sigma_{\theta}^4g(\mathbf{a})^2+\sigma_{\theta}^2}g(\mathbf{a})}\right)\;,
\end{align}where $(b)$ results since the unitary transformation $\mathbf{U}$ does not change the distribution of $\tilde{\mathbf{y}}$, and $(c)$ holds since  $\tilde{\mathbf{y}}^H\mathbf{G}\tilde{\mathbf{y}}$ has a scaled Chi-square distribution with two degrees of freedom. In a similar way, $P_{FA}$ can be derived as
\begin{equation}\label{eq:pf}
P_{FA}=\mathrm{exp}\left(-\frac{\gamma^{'}}{\sigma_{\theta}^2g(\mathbf{a})}\right)\;.
\end{equation}

\section{Neyman-Pearson Detector Optimization and Analysis}\label{sec:three}

Both $P_D$ and $P_{FA}$ are functions of the sensor transmission gains $\mathbf{a}$, and thus it is natural to find values for the entries of $\mathbf{a}$ that optimize detection performance.  In what follows we will show how to find $\mathbf{a}$ such that $P_D$ is maximized for a given $P_{FA}$.  According to (\ref{eq:pf}), the threshold required to achieve $P_{FA}=\epsilon$ is
\begin{equation}
\gamma'=-\sigma_\theta^2 g(\mathbf{a}) \ln{\epsilon}\;.
\end{equation}
When substituted into~(\ref{eq:pd2}), this threshold yields
\begin{equation}\label{eq:optpd}
P_{D}= \exp\left(\frac{\ln{\epsilon}}{\sigma_{\theta}^2g(\mathbf{a})+1}\right) \; .
\end{equation}
Since $\ln{\epsilon} < 0$, $P_{D}$ is maximized when the signal-to-noise ratio (SNR) $g(\mathbf{a})$ is maximized.  Thus, the problem becomes
\begin{eqnarray}\label{eq:minga}
\max_{\mathbf{a}}\!&&\!\!\!g(\mathbf{a})=\mathbf{a}^H\mathbf{H}^H\!(\mathbf{H}\mathbf{D}\mathbf{V}\mathbf{D}^H\mathbf{H}^H\!\!+\!\sigma_n^2\mathbf{I}_M)^{-1}\mathbf{H}\mathbf{a}\nonumber\\
s.t.\!&&\!\!\!\mathbf{a}^{H}\mathbf{a}=P\;,
\end{eqnarray}
where $P$ denotes the constraint on the sum sensor transmit power. This result was derived in \cite{Banavar:2012} by examining the behavior of the error exponent as the number of sensors went to infinity.  Here we see the result holds for fixed and finite values of $N$.  The role of $g(\mathbf{a})$ in determining estimation performance for $\theta$ has also been noted in \cite{Smith:2009,Jiang:20122}.  In general, finding a solution to~(\ref{eq:minga}) is difficult due to its nonlinear and non-convex dependence on $\mathbf{a}$.  A simpler solution was found to be possible in \cite{Jiang:20122} if the sensor gains were restricted to all have the same magnitude and only the phase was optimized.  In this case, the solution was shown to be found via a relaxed semi-definite program.  In this paper, we show that a closed-form ``water-filling'' type of solution for~(\ref{eq:minga}) is possible under the assumption that $M\rightarrow \infty$.

\subsection{Energy Efficiency}

For our analysis, we assume the wireless fading channel between the sensor node $i$ and FC is modeled as
\begin{equation}\label{eq:channel}
\mathbf{h}_i=\frac{\tilde{\mathbf{h}}_i}{\sqrt{d_i^{\alpha}}}\;,
\end{equation}
where $d_i$ is the distance between the sensor node and FC, $\alpha$ is the path loss exponent, and $\mathbf{\tilde{h}}_i\in\mathbb{C}^{M\times 1}$ is a complex Gaussian vector with distribution $\mathcal{CN}(\mathbf{0},\mathbf{I}_M)$. Note that the assumption here of independent and identically distributed channel coefficients is made primarily to enable the asymptotic analysis of the detection performance at the FC. The following theorem characterizes the energy efficiency of the NP detector for large $M$.

\begin{theorem}
Assuming Rayleigh fading wireless channels, as the number of FC antennas $M$ tends to infinity, the transmit gain $|a_i|^2$ at each sensor can be reduced by $1/M$ to \emph{almost surely} achieve the same optimal $P_D$ for a given fixed $P_{FA}$.
\end{theorem}
\begin{IEEEproof}
We will show that as $M\rightarrow \infty$, the function $g(\mathbf{a})$ in~\eqref{eq:optpd} and~\eqref{eq:minga} remains constant if the product $M |a_i|^2$ is held constant. We first use the matrix inversion lemma to show that
\begin{align}\label{eq:inv}
\left(\mathbf{H}\mathbf{D}\mathbf{V}\mathbf{D}^H\mathbf{H}^H+\sigma^2_n\mathbf{I}_M\right)^{-1}=\;\frac{1}{\sigma_n^2}\mathbf{I}_M-\frac{1}{\sigma_n^4}\mathbf{H}\left(\mathbf{E}^{-1}+\frac{1}{\sigma^2_n}\mathbf{H}^H\mathbf{H}\right)^{-1}\mathbf{H}^H\;,
\end{align}
where $\mathbf{E}=\mathbf{D}\mathbf{V}\mathbf{D}^H$. Note that we have assumed that $|a_i|>0$ to guarantee the matrix inverse $\mathbf{E}^{-1}$ exists, but we will see that the final solution allows $|a_i| \rightarrow 0$. Substituting (\ref{eq:inv}) into $g(\mathbf{a})$ yields
\begin{align}\label{eq:inv2}
g(\mathbf{a})=\frac{1}{\sigma_n^2}\mathbf{a}^H\mathbf{H}^H\mathbf{H}\mathbf{a}-\frac{1}{\sigma_n^4}\mathbf{a}^H\mathbf{H}^H\!\mathbf{H}\!\left(\!\mathbf{E}^{-1}\!\!+\!\!\frac{1}{\sigma^2_n}\mathbf{H}^H\mathbf{H}\!\right)^{\!\!\!-1}\!\!\!\mathbf{H}^H\mathbf{H}\mathbf{a}\;.
\end{align}
For large $M$, the product $\mathbf{H}^H\mathbf{H}$ converges almost surely to \cite{Ngo:2013}:
\begin{equation}\label{eq:approx}
\lim_{M\to\infty}\frac{1}{M}\mathbf{H}^H\mathbf{H}= \mathrm{diag}\left\{\frac{1}{d_1^{\alpha}}\cdots \frac{1}{d_N^{\alpha}}\right\}\; ,
\end{equation}
and substituting (\ref{eq:approx}) into (\ref{eq:inv2}) yields, after some calculations,
\begin{equation}\label{eq:approx2}
\lim_{M\to\infty} g(\mathbf{a})=\lim_{M\to\infty}\sum_{i=1}^N\frac{M|a_i|^2}{\sigma_n^2d_i^{\alpha}+
\sigma_{v,i}^2M|a_i|^2}\; .
\end{equation}
We see that $g(\mathbf{a})$ remains asymptotically unchanged as long as the product $M |a_i|^2$ is held constant, and thus asymptotically equivalent detection performance can be achieved if any decrease in sensor transmit power is balanced by a corresponding increase in the number of FC antennas.
\end{IEEEproof}

\subsection{Sensor Gain Optimization}

Based on (\ref{eq:approx2}), when $M \rightarrow \infty$, the original problem (\ref{eq:minga}) can be rewritten as
\begin{eqnarray}\label{eq:opt2}
\max_{|a_i|^2} &&\sum_{i=1}^N\frac{M|a_i|^2}{\sigma_n^2d_i^{\alpha}+\sigma_{v,i}^2M|a_i|^2}\\
 s.t. &&\sum_{i=1}^N |a_i|^2=P\;\label{eq:c2}.     \nonumber
\end{eqnarray}
We see from this formulation that as $M \rightarrow\infty$, only the magnitude of $a_i$ is important in determining the detection performance, and we see that there is no problem if $|a_i| \rightarrow 0$ for some $i$.  As $M$ grows, eventually we reach the point where $\sigma_{v,i}^2M|a_i|^2 \gg \sigma_n^2d_i^{\alpha}$, in which case the choice of the sensor gains no longer matters.  However, we will see in the simulations that for moderately large values of $M$, optimizing~(\ref{eq:opt2}) over $|a_i|$ provides a significant benefit, especially when $P$ is relatively small.

Define a new variable $x_i=|a_i|^2$, so that problem (\ref{eq:opt2}) is equivalent to
\begin{eqnarray}\label{eq:opt3}
\min_{x_i} &&\sum_{i=1}^N\frac{-Mx_i}{\sigma_n^2d_i^{\alpha}+\sigma_{v,i}^2M x_i}\\
 s.t. && \sum_{i=1}^N x_i=P\nonumber\\
      && 0\le x_i \;\nonumber.
\end{eqnarray}
In problem (\ref{eq:opt3}), the objective function is the sum of $N$ convex functions of $x_i$, and the constraints are linear with respect to the variable $x_i$, so~(\ref{eq:opt3}) is a convex problem and we can find a ``closed-form'' solution using the Karush-Kuhn-Tucker (KKT) conditions \cite{Boyd:2004}.

The Lagrangian of (\ref{eq:opt3}) is given by:
\begin{align}\label{eq:lagarange}
\mathcal{L}(x_i;\lambda,\mu_i)=&\sum_{i=1}^N\frac{-Mx_i}{\sigma_n^2d_i^{\alpha}+\sigma_{v,i}^2M x_i}+\lambda\left(\sum_{i=1}^{N}x_i-P\right)-\sum_{i=1}^N\mu_ix_i\;,
\end{align}
and the corresponding KKT conditions are as follows:
\begin{subequations}
\begin{align}
\frac{-\sigma_n^2d_i^{2\alpha}M}{(\sigma_n^2d_i^{\alpha}+\sigma_{v,i}^2M x_i)^2}+\lambda-\mu_i&=0\;\\
\lambda\left(\sum_{i=1}^Nx_i-P\right)&=0\;\\
\sum_{i=1}^Nx_i-P&= 0\;\\
x_i\mu_i&=0\;\label{eq:kkt3}\\
x_i,\; \mu_i,\;\lambda&\ge0\;.
\end{align}
\end{subequations}
After some simple manipulations, we arrive at the following optimal solution to~(\ref{eq:opt2}):
\begin{equation}\label{eq:optsol}
|a_{i}^*|=\sqrt{\frac{\left(\sqrt{\frac{\sigma_n^2d_i^{\alpha}M}{\lambda}}-\sigma_n^2d_i^{\alpha}\right)^{+}}{\sigma_{v,i}^2M}}\;,
\end{equation}
where $\lambda > 0$ is chosen such that $\sum_{i=1}^N |a_i^{*}|^2=P$. Lower and upper bounds for $\lambda$ are given by
\begin{subequations}
\begin{align}
\lambda_u&=\frac{M}{\sigma_n^2\min_i\{ d_i^{\alpha}\}}\\
\lambda_l&=\min_{i}\left\{\frac{\sigma_n^2d_i^{\alpha}M}
{(\sigma_n^2d_i^{\alpha}+\sigma_{v,i}^2PM)^2}\right\}\; ,
\end{align}
\end{subequations}
and the unique value of $\lambda$ can be found via a simple bisection search over $[\lambda_l, \lambda_u]$.

Note that while implementing the NP detector in~(\ref{eq:lemma1}) requires instantaneous CSI, the large $M$ assumption allows the optimal gains in~(\ref{eq:optsol}) to be computed using only the channel statistics, determined in this case by the distances of the FC to the sensors.  This is of interest since it means the sensors will not require frequent feedback from the FC to update their transmit gains.

\subsection{Single-Antenna FC}

It is of interest to consider the single-antenna FC case separately, both for purposes of comparison and because in this case an exact solution can be obtained.  When $M=1$, the signal model reduces to
\begin{subequations}
\begin{align}
\mathcal{H}_0: y&=\mathbf{a}^H\mathbf{F}\mathbf{v}+n \label{eq:rsigs1}\\
\mathcal{H}_1: y&=\mathbf{a}^H\mathbf{h}\theta+\mathbf{a}^H\mathbf{F}\mathbf{v}+n\;, \label{eq:rsigs2}
\end{align}
\end{subequations}
where $\mathbf{a}=[a_1\cdots a_N]^H$, $\mathbf{h}=[h_1 \cdots h_N]^T$, $\mathbf{F}=\mathrm{diag}\{h_1\cdots h_N\}$ and $h_i$ denotes the scalar channel gain between the $i$th sensor and the FC.
The conditional PDFs of $y$ under $\mathcal{H}_1$ and $\mathcal{H}_0$ are given by
\begin{subequations}
\begin{align}
p(y|\mathcal{H}_1)&=\frac{1}{\pi(\sigma_s^2+\sigma_w^2)} \mathrm{exp}\left(-\frac{|y|^2}{\sigma_s^2+\sigma_w^2}\right)\\
p(y|\mathcal{H}_0)&=\frac{1}{\pi\sigma_w^2}\mathrm{exp}\left(-\frac{|y|^2}{\sigma_w^2}\right)\;,
\end{align}
\end{subequations}
where $\sigma_s^2=\sigma_{\theta}^2\mathbf{a}^H\mathbf{h}\mathbf{h}^H\mathbf{a}$ and $\sigma_w^2=\mathbf{a}^H\mathbf{F}\mathbf{V}\mathbf{F}^H\mathbf{a}+\sigma^2_n$.

For a given threshold $\tilde{\gamma}$, the NP detector decides $\mathcal{H}_1$ if
\begin{equation}\label{eq:lrs}
L(y)=\frac{p(y|\mathcal{H}_1)}{p(y|\mathcal{H}_0)}>\tilde{\gamma}\;,
\end{equation}
which results in deciding $\mathcal{H}_1$ if
\begin{equation}\label{eq:singlenp}
|y|^2>\ln\left(\tilde{\gamma}\left(1+\frac{\sigma_s^2}{\sigma_w^2}\right)\right)\left(1+\frac{\sigma_w^2}{\sigma_s^2}\right)\sigma_w^2\;.
\end{equation}
Following an analysis similar to the multi-antenna case, the probability of detection $P_{D}^s$ and the probability of false alarm $P_{FA}^s$ for the single-antenna FC are given by
\begin{subequations}
\begin{align}
P_{D}^s&=\mathrm{exp}\left(-\frac{\tilde{\gamma}^{'}}{\sigma_s^2+\sigma_w^2}\right)\label{pfad} \\
P_{FA}^s&=\mathrm{exp}\left(-\frac{\tilde{\gamma}^{'}}{\sigma_w^2}\right)\label{pfas} \;,
\end{align}
\end{subequations}
where $\tilde{\gamma}^{'}= \ln\left(\tilde{\gamma}\left(1+\frac{\sigma_s^2}{\sigma_w^2}\right)\right)
\left(\sigma_w^2+\frac{\sigma_w^4}{\sigma_s^2}\right)$.

To fix $P_{FA}^s=\epsilon$, we set $\tilde{\gamma}^{'}=-\sigma_{w}^2\ln \epsilon$, and maximizing $P_{D}^s$ for a fixed $P_{FA}^s$ is equivalent to
\begin{eqnarray}\label{eq:opts}
\max_{\mathbf{a}} &&\frac{\sigma_s^2}{\sigma_w^2}=\frac{\sigma_{\theta}^2\mathbf{a}^H\mathbf{h}\mathbf{h}^H\mathbf{a}}
{\mathbf{a}^H\mathbf{F}\mathbf{V}\mathbf{F}^H\mathbf{a}+\sigma^2_n} \\
s.t. &&\mathbf{a}^{H}\mathbf{a}=P\; . \nonumber
\end{eqnarray}
Problem~(\ref{eq:opts}) is essentially identical to problem~(3) in \cite{Jiang:2013}, and using the same solution method derived in \cite{Jiang:2013} leads to
\begin{equation}\label{eq:optsols}
\tilde{\mathbf{a}}^{*}=\sqrt{\frac{P}{\mathbf{h}^H\mathbf{R}^{-2}\mathbf{h}}}\mathbf{R}^{-1}\mathbf{h}\;,
\end{equation}
where $\mathbf{R}=\mathbf{F}\mathbf{V}\mathbf{F}^H+\frac{\sigma^2_n}{P}\mathbf{I}_N$\;, and the maximum value of $\frac{\sigma_s^2}{\sigma_w^2}$ is
\begin{equation}\label{eq:optrs}
\left.\frac{\sigma_s^2}{\sigma_w^2}\right|_{\tilde{\mathbf{a}}^{*}}=\sigma_{\theta}^2\mathbf{h}^H\mathbf{R}^{-1}\mathbf{h}\;.
\end{equation}
In the following theorem, we compare the detection performance of single- and multi-antenna FCs under low and high transmit power scenarios.

\begin{theorem}\label{them:2}
Assume $P_{FA}=\epsilon$ and $M\to\infty$.  When $P=O(1/M) \rightarrow 0$, the NP detector implemented by an FC with $M$ antennas achieves a $P_D$ lower bounded by
\begin{equation}
P_D>\epsilon^{\frac{1}{1+\frac{\sigma_{\theta}^2}{3}\sum_{i=1}^N\frac{1}{\sigma_{v,i}^2}}}\;,\label{eq:lblm}
\end{equation}
while the $P_D^s$ for a single-antenna FC is bounded by
\begin{equation}
\epsilon < P_{D}^s< \epsilon^{\frac{1}{1+\zeta}},\label{eq:ubls}
\end{equation}
where $\zeta=\frac{1}{2M}\sum_{i=1}^N\frac{\sigma_{\theta}^2d_i^{\alpha}}{\sigma_{v,i}^2}\mathbf{h}^H\mathbf{h} \to 0$ in probability. When $P\to\infty$, both $P_D$ and $P_D^s$ converge from below to the same upper bound:
\begin{equation}
\left\{ P_D , P_D^s \right\} \uparrow \epsilon^{\frac{1}{1+\sigma_{\theta}^2\sum_{i=1}^N\frac{1}{\sigma_{v,i}^2}}} \; . \label{eq:ubh}
\end{equation}
\end{theorem}

\begin{IEEEproof}
See Appendix \ref{app:theorem2}\;.
\end{IEEEproof}

Theorem \ref{them:2} shows that when the transmit power $P$ goes to zero, $P_D^s$ for a single-antenna FC converges to $P_{FA}^s$ regardless of the sensor network scenario, while $P_D$ for a multi-antenna FC is strictly greater than $P_{FA}$, provided that $M \rightarrow\infty$ and $P\to 0$ no faster than $O(1/M)$.  When $\sigma_\theta^2$ is large and the $\sigma_{v,i}^2$ are small, $P_D$ can in fact still converge to a value near unity.  On the other hand, when $P$ is large, both $P_D$ and $P_D^s$ converge to the same upper bound, and there is no benefit to having multiple antennas at the FC.

\subsection{LMMSE Estimation}

While our paper is focused on detection, we show here that similar results hold for LMMSE estimation. According to the Gauss-Markov Theorem \cite{Kay:1993}, the LMMSE estimator of $\theta$ is
\begin{equation}\label{eq:lmsem}
\hat{\theta}=\frac{\mathbf{a}^H\mathbf{H}^H(\mathbf{H}\mathbf{D}\mathbf{V}\mathbf{D}^H\mathbf{H}^H\!+\!\sigma_n^2\mathbf{I}_M)^{-1}\mathbf{y}}{\sigma_{\theta}^{-2}\!+\!\mathbf{a}^H\mathbf{H}^H(\mathbf{H}\mathbf{D}\mathbf{V}\mathbf{D}^H\mathbf{H}^H\!+\!\sigma_n^2\mathbf{I}_M)^{-1}\mathbf{H}\mathbf{a}}\;,
\end{equation}
and the mean squared error is calculated as
\begin{align}\label{eq:mse}
\mathrm{MSE}(\hat{\theta})&=\mathbb{E}\{|\theta-\hat{\theta}|^2\}\nonumber\\
&=\frac{1}{\sigma_{\theta}^{-2}+ g(\mathbf{a})}\;,
\end{align}
where $g(\mathbf{a})=\mathbf{a}^H\mathbf{H}^H(\mathbf{H}\mathbf{D}\mathbf{V}\mathbf{D}^H\mathbf{H}^H+\sigma_n^2\mathbf{I}_M)^{-1}\mathbf{H}\mathbf{a}$, as defined in~(\ref{eq:minga}). Thus, the problem of choosing the gains $\mathbf{a}$ to minimize the MSE is identical to the problem of maximizing $P_D$ for a fixed $P_{FA}$ in~(\ref{eq:minga}), and the same conclusions drawn above regarding energy efficiency and the optimal sensor gains apply here as well.  This is also true for the single-antenna FC, as it can be easily shown that minimizing MSE requires maximization of $\sigma_s^2/\sigma_w^2$, as with the NP detector.

The following corollary to Theorem~\ref{them:2} can also be established.
\begin{corollary}
When $M\to \infty$ and $P=O(1/M) \rightarrow 0$, the MSE of the LMMSE estimator of $\theta$ is upper bounded by
\begin{equation}\label{eq:mseub}
\mathrm{MSE}(\hat{\theta})<\frac{1}{\sigma_{\theta}^{-2}+\frac{1}{3}\sum_{i=1}^N\frac{1}{\sigma_{v,i}^2}}\;,
\end{equation}
while the MSE achieved by the single-antenna FC is bounded by
\begin{equation}
\frac{\sigma_{\theta}^2}{1+\zeta}<\mathrm{MSE}(\hat{\theta}_s)<\sigma_{\theta}^2\;,\label{eq:mselb}
\end{equation}
where $\zeta=\frac{1}{2M}\sum_{i=1}^N\frac{\sigma_{\theta}^2d_i^{\alpha}}{\sigma_{v,i}^2}\mathbf{h}^H\mathbf{h} \to 0$ in probability. When $P\to\infty$, both MSEs converge from above to the same lower bound:
\begin{equation}
\mathrm{MSE}(\hat{\theta},\hat{\theta}_s) \ge \frac{1}{\sigma_{\theta}^{-2}+\sum_{i=1}^N\frac{1}{\sigma_{v,i}^2}}\label{eq:msehm} \; .
\end{equation}
\end{corollary}
\begin{IEEEproof}
The proof essentially follows that for Theorem~\ref{them:2} and is thus omitted.
\end{IEEEproof}

\section{Energy Detector Analysis and Sensor Gain Optimization}\label{sec:four}

Obtaining the instantaneous CSI required for the NP detector consumes sensor power and could be difficult in fast fading scenarios.  Computing the NP test statistic also requires the inverse of the $M\times M$ channel-dependent matrix $\mathbf{C}_w$, which may be challenging when $M$ is large. Consequently, it is of interest to study computationally simpler approaches for detection in sensor networks that can be applied when the CSI for the sensors is unknown. In this section, we examine the performance of the energy detector (ED), which decides $\mathcal{H}_1$ if
\begin{equation}\label{eq:eddef}
T=\frac{1}{M}\mathbf{y}^H\mathbf{y}>\hat{\gamma}\;,
\end{equation}
for some predefined threshold $\hat{\gamma}$.

Under either $\mathcal{H}_0$ or $\mathcal{H}_1$, the test statistic $T$ can be expressed as
\begin{equation}\label{eq:eddef2}
T=\frac{1}{M}\sum_{i=1}^M\frac{\lambda_i}{2}\chi_{i}^2(2)\;,
\end{equation}
where $\lambda_i$ is the $i$th eigenvalue of the covariance matrix $\mathbf{C}_w$ (under $\mathcal{H}_0$) or $\mathbf{C}_s+\mathbf{C}_w$ (under $\mathcal{H}_1$) and the $\chi_{i}^2(2)$ terms represent independent chi-squared random variables with two degrees of freedom.  Thus, while the ED test statistic does not require CSI, computing the ED probability of detection $P_D^e$ and false alarm $P_{FA}^e$ does. When $M$ is large, one could consider approximating $T$ as a normal random variable using the Central Limit Theorem.  However, because the largest $N$ eigenvalues of $\lambda_i$ will increase with $M$, Lindeberg's condition is not satisfied and the normal distribution can not provide a good approximation for $T$. Even if the distribution of $T$ could be computed, it would be a complicated function of the transmit gains $\mathbf{a}$ and would be difficult to optimize. Instead, in the following we will use the so-called {\em deflection} \cite{Kay:19932, Picinbono:1995, Unnikrishnan:2008, Quan:2008} of $T$ as the metric of detection performance, which will allow us to obtain an optimal value for $\mathbf{a}$ that does not depend on CSI as $M \to \infty$.

\subsection{Energy Efficiency}

The deflection coefficient for a given test statistic $T$ is defined as \cite{Kay:19932}
\begin{equation}\label{eq:deflec}
D(T)=\frac{\left(\mathbb{E}\{T|\mathcal{H}_1\}-
\mathbb{E}\{T|\mathcal{H}_0\}\right)^2}{\mathrm{Var}\{T|\mathcal{H}_0\}}\;.
\end{equation}
The deflection metric in (\ref{eq:deflec}) can be viewed as the normalized distance between the distributions of $T$ under $\mathcal{H}_0$ or $\mathcal{H}_1$, and is generally regarded as an accurate metric for characterizing detection performance \cite{Picinbono:1995}. Note that a \emph{modified deflection} is proposed in \cite{Quan:2008}, which replaces $\mathrm{Var}\{T|\mathcal{H}_0\}$ in (\ref{eq:deflec}) with $\mathrm{Var}\{T|\mathcal{H}_1\}$.  As mentioned below, both deflection statistics yield very similar problem formulations that can be solved via the same approach. As derived in the following theorem, one of the key properties of the energy detector for our WSN application is that the sensor transmit power can be reduced by a factor of $1/\sqrt{M}$ to maintain a constant deflection as $M \to \infty$.

\begin{theorem}\label{the:constdefl}
Assuming Rayleigh fading channels, the deflection of the test statistic $T=\frac{1}{M}\mathbf{y}^H\mathbf{y}$ \emph{almost surely} remains constant as $M \to \infty$ provided that the sensor transmit power satisfies $|a_i|^2=\frac{P_i}{\sqrt{M}}$ for arbitrary constant $P_i$\;.
\begin{IEEEproof}
See Appendix \ref{app:theorem3}\;.
\end{IEEEproof}
\end{theorem}

\subsection{Sensor Gain Optimization}

As with the NP detector, the proof of Theorem~\ref{the:constdefl} shows that as $M \to \infty$, only the magnitude $|a_i|$ of the sensor transmission gains influences the deflection. In this section, we address the problem of finding the $|a_i|$ that maximize the deflection under a sum power constraint. The power allocation problem is formulated as
\begin{eqnarray}\label{eq:opte1}
\max_{|a_i|^2} && D\left(T\right)\\
 s.t. &&\mathbf{a}^H\mathbf{a}=P\;.\nonumber
\end{eqnarray}
According to (\ref{eq:deflec2}), we can rewrite~(\ref{eq:opte1}) as
\begin{eqnarray}\label{eq:opte2}
\max_{x_i} && \frac{\mathbf{x}^T\mathbf{d}\mathbf{d}^T\mathbf{x}}{\mathbf{x}^T\mathbf{B}\mathbf{x}+
\frac{2\sigma_n^2}{M}\mathbf{b}^T\mathbf{x}+\frac{\sigma_n^4}{M}}\\
 s.t. &&\mathbf{e}^T\mathbf{x}= P\label{eq:cons}\nonumber\\
  &&0\le x_i\;,\;i=1,\cdots,N\;,\nonumber
\end{eqnarray}
where
\begin{subequations}
\begin{align}
\mathbf{x}&=[|a_1|^2 \; \cdots \; |a_N|^2]^T \label{eq:d1}\\
\mathbf{d}&=\left[\frac{1}{d_1^{\alpha}} \; \cdots \; \frac{1}{d_N^{\alpha}}\right]^T \\
\mathbf{B}&=\mathrm{diag}\left\{\frac{\sigma_{v,1}^4}{d_1^{2\alpha}} \; \cdots \;
\frac{\sigma_{v,N}^4}{d_N^{2\alpha}}\right\} \\
\mathbf{b}&=\left[\frac{\sigma_{v,1}^2}{d_1^{\alpha}} \; \cdots \; \frac{\sigma_{v,N}^2}{d_N^{\alpha}} \right]^T \label{eq:d2}\\
\mathbf{e}&=[1 \; \cdots \; 1]^T \;.
\end{align}
\end{subequations}
We note here that if the modified deflection of \cite{Quan:2008} is used instead, then the resulting problem is identical to~(\ref{eq:opte2}), except for the definitions of $\mathbf{B}$ and $\mathbf{b}$, which become
\begin{subequations}
\begin{align}
\mathbf{B}'&=\mathrm{diag}\left\{\frac{\sigma_{v,1}^4+\sigma_{v,1}^2\sigma_\theta^2}{d_1^{2\alpha}} \; \cdots \; \frac{\sigma_{v,N}^4+\sigma_{v,N}^2\sigma_\theta^2}{d_N^{2\alpha}}\right\} \\
\mathbf{b}'&=\left[\frac{\sigma_{v,1}^2+\sigma_\theta^2}{d_1^{\alpha}} \; \cdots \; \frac{\sigma_{v,N}^2+\sigma_\theta^2}{d_N^{\alpha}} \right]^T \;.
\end{align}
\end{subequations}
Thus, the solution to~(\ref{eq:opte2}) described below can be applied directly to the modified deflection as well.

Problem (\ref{eq:opte2}) is the maximization of the ratio of two quadratic functions under quadratic constraints, which is referred to as a QCRQ problem. In \cite{Beck:2010}, a solution to the QCRQ problem is found by converting it to a semidefinite program (SDP) via rank relaxation, followed by an eigendecomposition to find a rank-one result.  However, in general, the optimality of the rank-one solution to the original problem can not be guaranteed.  Consequently, here we take a different approach and find an asymptotically optimal solution by maximizing an upper bound for (\ref{eq:opte2}) that is tight when $M\to\infty$. In particular, we consider
\begin{eqnarray}\label{eq:opte3}
\max_{x_i} && \frac{\mathbf{x}^T\mathbf{d}\mathbf{d}^T\mathbf{x}}{\mathbf{x}^T\mathbf{B}\mathbf{x}+\frac{\sigma_n^4}{M}}\\
 s.t. &&\mathbf{e}^T\mathbf{x}= P\label{eq:cons2}\nonumber\\
  &&0\le x_i\;,\;i=1,\cdots,N\;.\nonumber
\end{eqnarray}
It is easy to verify that (\ref{eq:opte3}) provides an upper bound for (\ref{eq:opte2}) and that the bound is asymptotically achieved when $M\to\infty$.  Since $M\to \infty$, we could eliminate the second term in the denominator of~(\ref{eq:opte3}) as well, but we will see in the simulations that it is advantageous to keep it, especially in situations where $P$ is small.  The simplification that arises when this term is dropped will be discussed later, when asymptotic solutions for large $P$ are investigated. In the following, we will show that~(\ref{eq:opte3}) can be converted to a  quadratically constrained linear program (QCLP) \cite{Martein:1987} and solved via standard convex optimization methods.

First, we rewrite~(\ref{eq:opte3}) as
\begin{subequations}
\begin{eqnarray}\label{eq:opte4}
\max_{x_i} && \frac{\mathbf{x}^T\mathbf{d}\mathbf{d}^T\mathbf{x}}{\mathbf{x}^T\tilde{\mathbf{B}}\mathbf{x}}\\
 s.t. &&\mathbf{e}^T\mathbf{x}=P\label{eq:cons3}\\
  &&0\le x_i\;,\;i=1,\cdots,N\;,\nonumber
\end{eqnarray}
\end{subequations}
where $\tilde{\mathbf{B}}=\mathbf{B}+\frac{\sigma_n^4}{MP^2}\mathbf{e}\mathbf{e}^T$. Since the objective function in~(\ref{eq:opte4}) is unchanged by a simple scaling of $\mathbf{x}$, we do not need to explicitly consider the constraint in~(\ref{eq:cons3}) in maximizing~(\ref{eq:opte4}), and the optimal solution can be found via the following two steps:
\begin{enumerate}
\item Solve
\begin{eqnarray}\label{eq:opte5}
\max_{x_i} && \frac{\mathbf{x}^T\mathbf{d}\mathbf{d}^T\mathbf{x}}{\mathbf{x}^T\tilde{\mathbf{B}}\mathbf{x}}\\
 s.t.  &&0\le x_i\;,\;i=1,\cdots,N\;.\nonumber
\end{eqnarray}
\item Denote the result of (\ref{eq:opte5}) as $\tilde{\mathbf{x}}^{*}$, then the optimal solution to~(\ref{eq:opte4}) is given by
\begin{equation}\label{eq:scale}
\mathbf{x}^{*}=\frac{1}{\mathbf{e}^T\tilde{\mathbf{x}}^{*}}\tilde{\mathbf{x}}^{*}\;.
\end{equation}
\end{enumerate}

To solve problem (\ref{eq:opte5}), we first rewrite it in the equivalent form
\begin{subequations}
\begin{eqnarray}\label{eq:opte6}
\max_{x_i} && \mathbf{x}^T\mathbf{d}\mathbf{d}^T\mathbf{x}\\
 s.t.  &&\mathbf{x}^T\tilde{\mathbf{B}}\mathbf{x}=1\label{eq:cons4}\\
 &&0\le x_i\;,\;i=1,\cdots,N\;.\nonumber
\end{eqnarray}
\end{subequations}
To convert~(\ref{eq:opte6}) to a  QCLP, we make the following two observations: (1) since the elements of $\mathbf{x}$ and $\mathbf{d}$ are non-negative, maximizing $\mathbf{x}^T\mathbf{d}\mathbf{d}^T\mathbf{x}$ is equivalent to maximizing $\mathbf{x}^T\mathbf{d}$, and (2) we can relax the equality constraint in~(\ref{eq:cons4}) to an inequality $\mathbf{x}^T\tilde{\mathbf{B}}\mathbf{x} \le 1$, since we can always increase the objective function in~(\ref{eq:opte6}) by scaling $\mathbf{x}$ up to meet the constraint with equality.  Thus, solving~(\ref{eq:opte5}) is equivalent to solving the QCLP
\begin{eqnarray}\label{eq:opte9}
\min_{x_i} && -\mathbf{x}^T\mathbf{d}\\
 s.t. &&\mathbf{x}^T\tilde{\mathbf{B}}\mathbf{x}\le 1\nonumber\\
  &&0\le x_i\;,\;i=1,\cdots,N\; , \nonumber
\end{eqnarray}
for which straightforward convex optimization methods exist.  The final result for the original problem in~(\ref{eq:opte3}) is found by scaling the optimal solution to~(\ref{eq:opte9}) according to~(\ref{eq:scale}) to satisfy the power constraint.

Our simulation results in Section~\ref{sec:six} validate the use of the deflection to optimize detection performance.  In particular, we will see that performance improves as the deflection is increased and that with the $a_i$ chosen to maximize the deflection, detection performance remains asymptotically constant as $M \to \infty$ if the power constraint $P$ is scaled by $1/\sqrt{M}$.

\subsection{Single-Antenna FC}

For comparison purposes, we derive the deflection for the case of a single-antenna FC. Based on the signal model in equations~(\ref{eq:rsigs1}) and (\ref{eq:rsigs2}), the single-antenna deflection is given by
\begin{align}
D(T_s)&=\frac{\left(\mathbb{E}\{T_s|\mathcal{H}_1\}-
\mathbb{E}\{T_s|\mathcal{H}_0\}\right)^2}{\mathrm{Var}\{T_s|\mathcal{H}_0\}}\nonumber\\
&=\left(\frac{\sigma_{\theta}^2\mathbf{a}^H\mathbf{h}\mathbf{h}^H\mathbf{a}}
{\mathbf{a}^H\mathbf{F}\mathbf{V}\mathbf{F}^H\mathbf{a}+\sigma^2_n}\right)^2\;, \label{eq:c49}
\end{align}
where $T_s=|y|^2$ and $y$, $\mathbf{a}$, $\mathbf{h}$ and $\mathbf{F}$ are as defined in equation~(\ref{eq:rsigs2}). Unlike the deflection in~(\ref{eq:opte2}) when $M\rightarrow\infty$, it is easy to verify that $D(T_s)$ in~(\ref{eq:c49}) decreases monotonically as the norm of the transmission gain $\mathbf{a}$ decreases.  If channel state information is available at the FC, then the optimal gains that maximize $D(T_s)$ are given by~(\ref{eq:optsols}). A different approach is required in the single-antenna case without CSI; for example, in the simulations later we assume the sensor nodes transmit with equal power.  We will also observe in the simulation results that when the sum transmission power decreases, the probability of detection for the single-antenna FC will decrease accordingly, while the performance of the multi-antenna FC remains constant as long as the number of antennas increases proportionally to the square of the power decrease.

\subsection{Asymptotic Closed-form Solutions}

While convergence to a globally optimal solution is guaranteed for the  QCLP problem described above, we show here that direct closed-form solutions can be found for low and high SNR scenarios $P \gg \sigma_n^2$ and $P \ll \sigma_n^2$. When $P\gg \sigma_n^2$, the size of $\mathbf{x}^T\mathbf{B}\mathbf{x}$ in the denominator of the objective function~(\ref{eq:opte2}) will dominate the terms involving $M$, which are already small for large $M$.  Thus, for $P \gg \sigma_n^2$, another upper bound for~(\ref{eq:opte2}) is given by
\begin{equation}\label{eq:edub}
\frac{\mathbf{x}^T\mathbf{d}\mathbf{d}^T\mathbf{x}}{\mathbf{x}^T\mathbf{B}\mathbf{x}+
2\frac{\sigma_n^2}{M}\mathbf{b}^T\mathbf{x}+\frac{\sigma_n^4}{M}}<
\frac{\mathbf{x}^T\mathbf{d}\mathbf{d}^T\mathbf{x}}
{\mathbf{x}^T\mathbf{B}\mathbf{x}}\;.
\end{equation}
We can formulate the problem of maximizing this upper bound as
\begin{eqnarray}
\max_{x_i} && \frac{\mathbf{x}^T\mathbf{d}\mathbf{d}^T\mathbf{x}}{\mathbf{x}^T\mathbf{B}\mathbf{x}}\\
 s.t. &&\mathbf{e}^T\mathbf{x}=P\nonumber\\
  &&0\le x_i\;,\;i=1,\cdots,N\;,\nonumber
\end{eqnarray}
which has a closed-form solution since $\mathbf{B}$ and $\mathbf{d}$ have non-negative elements:
\begin{equation}
\mathbf{x}^{*}=\frac{P}{\mathbf{e}^T\mathbf{B}^{-1}\mathbf{d}}\mathbf{B}^{-1}\mathbf{d}\;,
\end{equation}
and the corresponding $a_i$ are
\begin{equation}\label{eq:clssol}
|a_i|=\sqrt{\frac{P}{\sum_{i=1}^N\frac{d_i^{\alpha}}{\sigma_{v,i}^4}}}\frac{d_i^\frac{\alpha}{2}}{\sigma_{v,i}^2}\;.
\end{equation}
Thus, for high SNR, after normalizing for distance, the sensors with the lowest measurement noise are allocated higher power.

When $P\ll \sigma_n^2$, the terms involving $\mathbf{x}$ in the denominator of~(\ref{eq:opte2}) will decrease faster than $1/M$, and thus the term $\frac{\sigma_n^2}{M}$ will eventually dominate.  This leads to the simpler optimization problem
\begin{eqnarray}\label{eq:opte10}
\max_{x_i} && \mathbf{x}^T\mathbf{d}\mathbf{d}^T\mathbf{x}\\
 s.t. &&\mathbf{e}^T\mathbf{x}=P\nonumber\\
  &&0\le x_i\;,\;i=1,\cdots,N\;.\nonumber
\end{eqnarray}
This is equivalent to maximizing the weighted sum $\mathbf{x}^T\mathbf{d}$ with constraint $\mathbf{e}^T\mathbf{x}=P$, and the optimal solution is to simply allocate all of the power to the sensor that is closest to the FC:
\begin{equation}\label{eq:clssol2}
|a_i|=\left\{\begin{array}{cl}
\sqrt{P}\; & \textrm{$i=\arg\min_i d_i$}\\
0\;  & \textrm{otherwise}\;.
\end{array} \right.
\end{equation}
Later in the simulation results, we will show that the solutions in~(\ref{eq:clssol}) and~(\ref{eq:clssol2}) provide good approximations to the optimal solution of problem~(\ref{eq:opte9}) for very large and very small values of the available sum power $P$, respectively.

\subsection{Detection Threshold Calculation}
Once the transmission gains $a_i$ of the sensor nodes are optimized, we need to find the threshold $\hat{\gamma}$ to achieve the desired PFA. In the following, we will show that asymptotically as $M\rightarrow\infty$, the value of $\hat{\gamma}$ can be calculated according to (\ref{eq:eddef2}) without requiring CSI. Under $\mathcal{H}_0$, the eigenvalues of $\mathbf{C}_w$ are given by
\begin{equation}\label{eq:eignh0}
\lim_{M\to\infty}\lambda_i\{\mathbf{C}_w\}=\left\{\begin{array}{lc} M\eta_i+\sigma_n^2 & \textrm{$1\le i \le N$}\\
\sigma_n^2 &\textrm{$N<i\le M$}\;,
\end{array} \right.
\end{equation}
where $\eta_i=\frac{|a_i|^2\sigma_{v,i}^2}{d_i^{\alpha}}$\;. Substituting~(\ref{eq:eignh0}) into (\ref{eq:eddef2}), we have
\begin{equation}\label{eq:edtest}
\lim_{M\to\infty}T=\sum_{i=1}^N\frac{1}{2}\left(\eta_i+\frac{\sigma_n^2}{M}\right)
\chi_{i}^2(2)+\frac{\sigma_n^2}{2M}\sum_{i=N+1}^M\chi_{i}^2(2) \; .
\end{equation}
According to the Strong Law of Large Numbers,
\begin{equation}
\lim_{M\to\infty}\frac{\sigma_n^2}{2M}\sum_{i=N+1}^M\chi_{i}^2(2)=\frac{M-N}{M}\sigma_n^2\;,
\end{equation}
and this equation holds almost surely. Then the right hand side of equation (\ref{eq:edtest}) can be viewed as the sum of weighted chi-square variables plus a constant, and for a specific $\hat{\gamma}$, the PFA is calculated as
\begin{align}\label{eq:edthrld}
P_{FA}&=\mathrm{Pr}\left(\lim_{M\to\infty}T>\hat{\gamma}|\mathcal{H}_0\right)\nonumber\\
&=\mathrm{Pr}\left(\sum_{i=1}^N\frac{1}{2}\left(\eta_i+\frac{\sigma_n^2}{M}\right)\chi_{i}^2(2)>\hat{\gamma}-\frac{M-N}{M}\sigma_n^2\right)\nonumber\\
&\overset{(k)}{=}\sum_{i=1}^N\frac{\left(\eta_i+\frac{\sigma_n^2}{M}\right)^{N-1}}{\prod_{l\neq i}(\eta_i-\eta_l)}e^{-\frac{1}{\eta_i+\frac{\sigma_n^2}{M}}\left(\hat{\gamma}-\frac{M-N}{M}\sigma_n^2\right)}\;,
\end{align}
where in $(k)$ we used a result from \cite{Al_Naffouri:2009}, and we assume that the values of $\eta_i$ are distinct. In the limit the PFA expression is independent of the CSI, and the value of the threshold $\hat{\gamma}$ that achieves the desired PFA can be found numerically using~(\ref{eq:edthrld}).

\section{Simulation Results}\label{sec:six}

In the simulation examples that follow, we assume $\sigma_{\theta}^2=1, \sigma_n^2=0.3, \alpha=2$ and $N=10$ sensor nodes.  The distances $d_i$ were uniformly distributed over $[2, 10]$, and the measurement noise powers $\sigma_{v,i}^2$ were uniformly distributed in the interval $[0.25, 0.5]$.  Once generated, $d_i$ and $\sigma_{v,i}^2$ were held fixed for all simulations.  Each point in the following plots is the result of averaging over 10000 trials for each of 300 scenarios; each trial involved a new random parameter $\theta$, as well as new noise realizations and each scenario has a new channel. Plots showing probability of detection were computed assuming a false alarm probability of $\epsilon=0.05$.  For the energy detector, both the deflection and modified deflection gave essentially the same performance, so only the results for the deflection are included.

\begin{figure}[!htb]
\centering
\includegraphics[height=3.5in, width=4.5in]{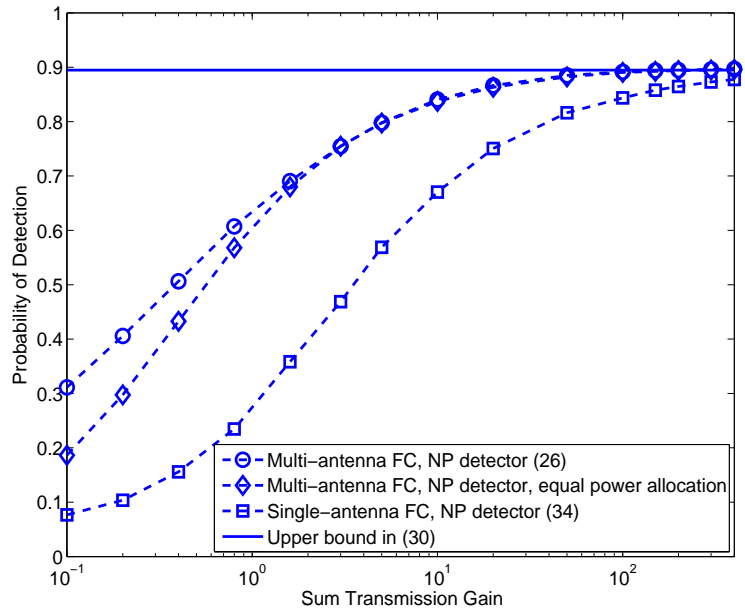}
\caption{Probability of detection for NP detector  vs. the value of $P$, with antenna number $M=50$.}\label{f1}
\end{figure}

\begin{figure}[!htb]
\centering
\includegraphics[height=3.5in, width=4.5in]{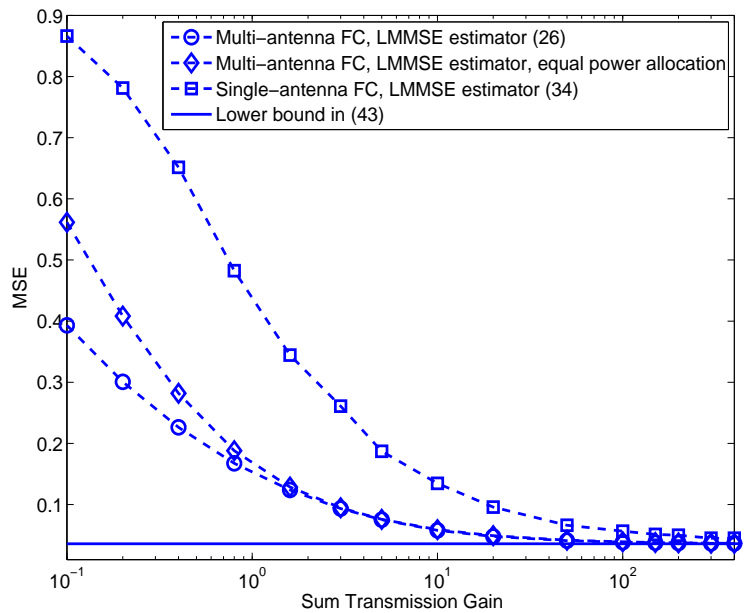}
\caption{Mean squared error vs. the value of $P$, with number of antennas $M=50$.}\label{f2}
\end{figure}

Figs.~\ref{f1} and~\ref{f2} show the NP detection and LMMSE estimation performance for a single-antenna FC and a multi-antenna FC with $M=50$ as the available power $P$ ranges from 0.1 to 400. As predicted, as $P$ grows, the performance benefit of having multiple antennas at the FC is eventually lost, with both curves in Fig.~\ref{f1} approaching the upper bound in~(\ref{eq:ubh}) and both curves in Fig.~\ref{f2} approaching the lower bound in~(\ref{eq:msehm}). However, in both cases the bound is reached with a much smaller value of $P$ in the multi-antenna case. Note also that for the multi-antenna FC, use of the optimal sensor transmit gains can achieve significantly better performance than equal power allocation when the sum transmit power is low.

\begin{figure}[!htb]
\centering
\includegraphics[height=3.5in, width=4.5in]{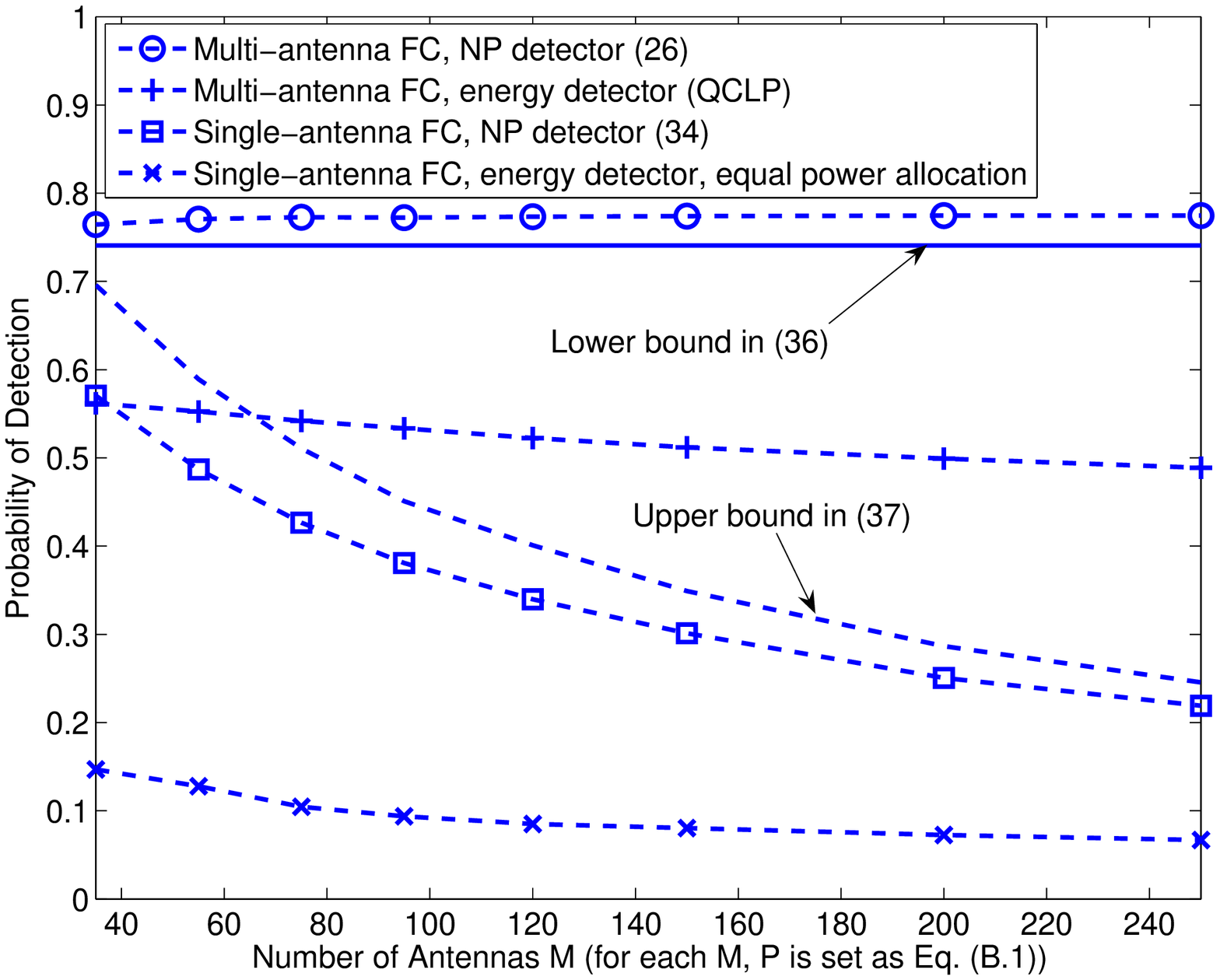}
\caption{Probability of detection vs. number of antennas $M$.}\label{f3}
\end{figure}

\begin{figure}[!htb]
\centering
\includegraphics[height=3.5in, width=4.5in]{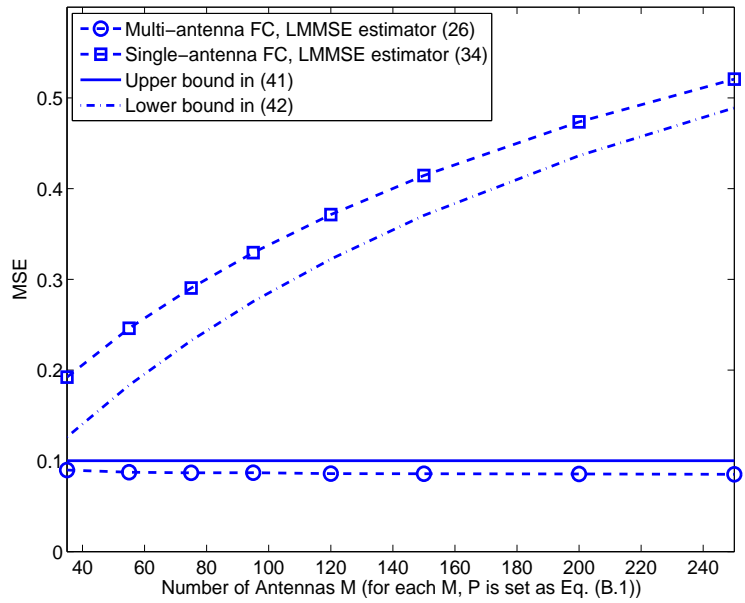}
\caption{Mean squared error vs. number of antennas $M$.}\label{f4}
\end{figure}

Figs.~\ref{f3} and~\ref{f4} respectively present the detection and estimation performance of single- and multi-antenna FCs for increasing $M$, with the sum power decreasing as $O(1/M)$ according to the formula $P=\sum_{i=1}^N \frac{\sigma_n^2d_i^{\alpha}}{2\sigma_{v,i}^2M}$. The energy efficiency of the multi-antenna NP detector is evident, as the MSE and $P_D$ are unchanged as $M$ increases and $P$ decreases; however, the performance of the multi-antenna ED detector degrades with $M$ as the sum power is decreasing at a rate faster than $1/\sqrt{M}$.  The lower bound in~(\ref{eq:lblm}) and the upper bound in~(\ref{eq:mseub}) provide tight estimates of the multi-antenna NP probability of detection and LMMSE estimation error, respectively.  The value of choosing the optimal sensor gains is evident in comparing the two detection curves for the single-antenna FC, which show a large gap in performance between that achieved with the optimal gains and simply assigning equal gains to all sensors.  The latter approach provides a $P_D^s$ that is barely greater than $P_{FA}^s$, while the optimal sensor gains have much better performance, although $P_D^s$ is decreasing due to the reduction in power.  The single-antenna upper bound in~(\ref{eq:ubls}) grows tight as $M$ increases, and is decreasing towards the lower bound $\epsilon$, albeit very slowly.

\begin{figure}[!htb]
\centering
\includegraphics[height=3.5in, width=4.5in]{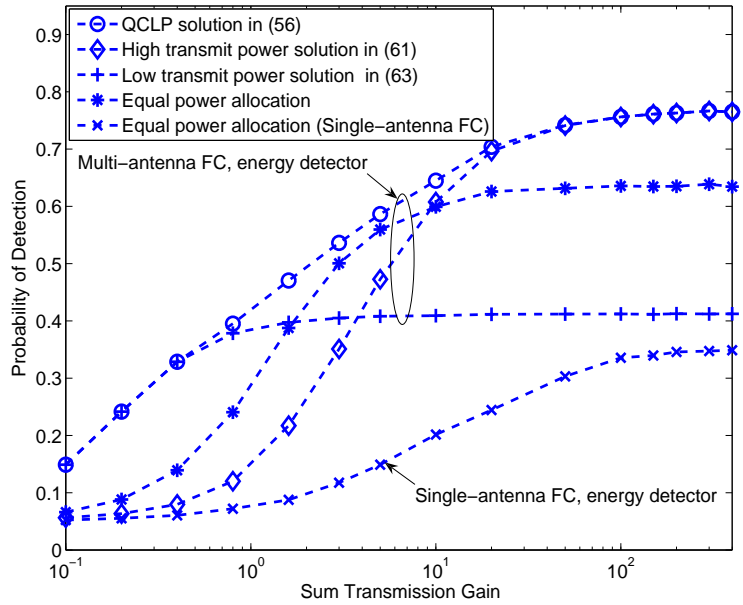}
\caption{Probability of detection for energy detector vs. the value of $P$, with number of antennas $M=50$.}\label{f5}
\end{figure}

\begin{figure}[!htb]
\centering
\includegraphics[height=3.5in, width=4.5in]{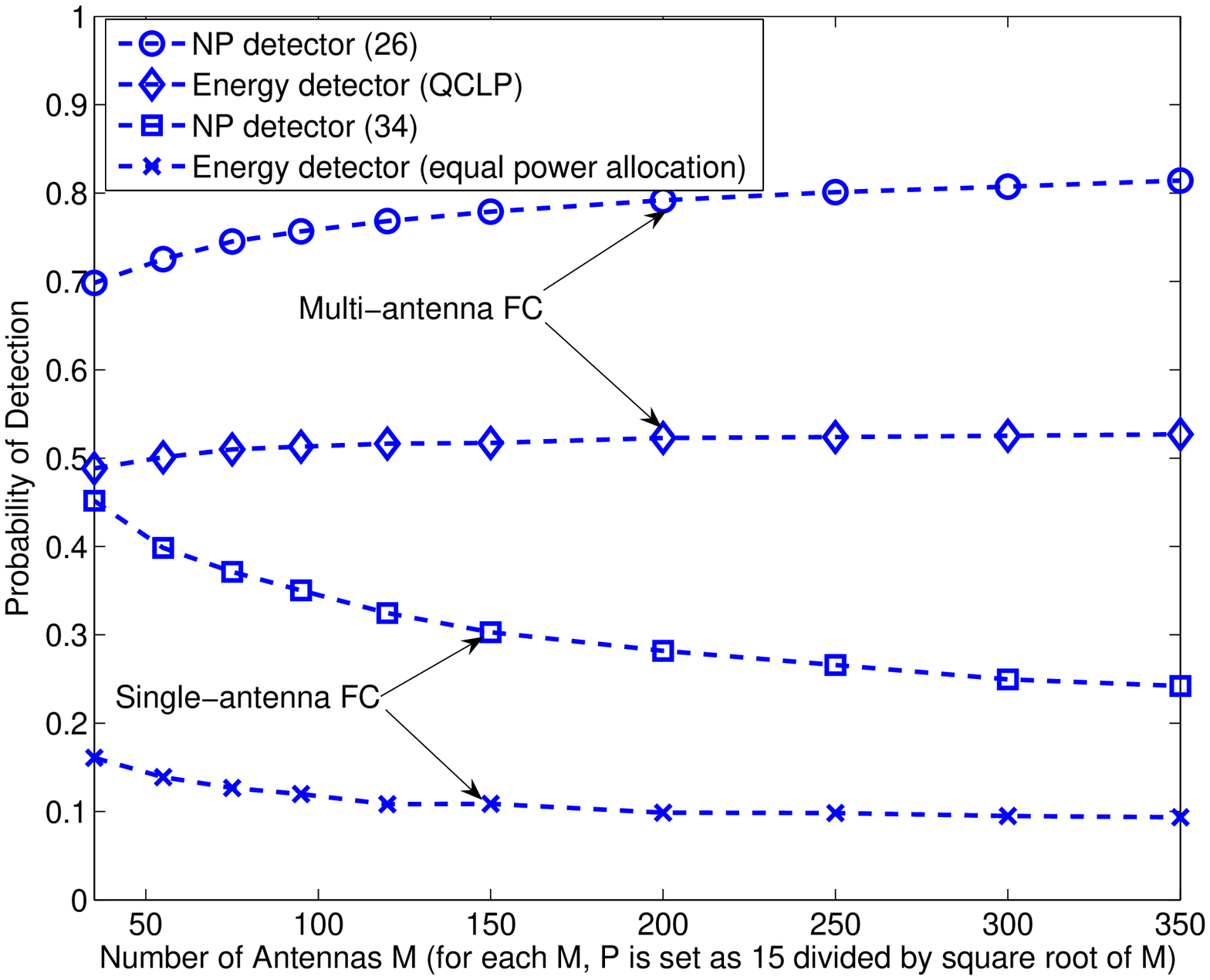}
\caption{Probability of detection vs. number of antennas $M$.}\label{f6}
\end{figure}

Fig.~\ref{f5} illustrates the detection performance of the ED approach with $P$ varying from $0.1$ to $400$.  The optimal QCLP approach is plotted along with the low and high SNR approximations in~(\ref{eq:clssol}) and~(\ref{eq:clssol2}), the ED implemented with equal power allocation to all sensors, and the single-antenna FC.  The low SNR approximation matches the QCLP approach for $P \le 1$, while the high SNR solution is optimal for $P \ge 20$; in between these values, the QCLP algorithm provides significantly better performance, although the equal power allocation is close for some values of $P$.  Unlike the NP detector, the single- and multiple-antenna ED solutions do not converge to the same performance for large $P$; we see in this example that there is a large performance benefit in having a multi-antenna FC, even for large $P$. In Fig.~\ref{f6}, we compare NP and energy detection performance as a function of $M$ assuming that $P=15/\sqrt{M}$.  Consistent with our analytical predictions, the ED with sensor gains chosen via the QCLP to maximize deflection has constant $P_D$, while the multi-antenna NP detector slowly improves and the single-antenna FC solutions degrade as $M$ increases.
\section{Conclusion}\label{sec:seven}
We have studied the detection and estimation performance of a sensor network communicating over a coherent multiple access channel with a fusion center possessing a large number of antennas.  We studied Neyman-Pearson and energy detection, derived optimal sensor transmission gains for each case, and showed that the optimal gains are phase-independent as the number of antennas grows large.  Similar to properties of massive MIMO wireless communications, one can trade antennas at the fusion center for energy efficiency at the sensors.  For the case of Neyman-Pearson detection and LMMSE estimation, which require channel state information, constant levels of performance can be achieved if the transmit power at the sensors is reduced proportional to the gain in the number of antennas.  For energy detection, which does not require channel state information, a constant deflection coefficient can be maintained if power is reduced proportional to the inverse square root of the number of antennas.  While bounds derived for Neyman-Pearson detection and LMMSE estimation show performance gains for a multiple-antenna fusion center in low sensor transmit power scenarios, the benefit is shown to disappear when the transmit power is high.  However, for the energy detector, having multiple antennas at the fusion center provides a significant advantage even when the sensors have high power.

\appendices
\numberwithin{equation}{section}
\section{Proof of Lemma 1}\label{app:lemma1}
Substituting $p(\mathbf{y}|\mathcal{H}_1)$ and $p(\mathbf{y}|\mathcal{H}_0)$ from~(\ref{eq:pdf1}) and (\ref{eq:pdf2}) into~(\ref{eq:lr}) and calculating the logarithm of (\ref{eq:lr}), we have
\begin{equation}\label{eq:loglr}
\mathbf{y}^H(\mathbf{C}_w^{-1}-\left(\mathbf{C}_s\!+\!\mathbf{C}_w\right)^{-1})\mathbf{y}>\ln\left(\gamma(1\!+\!\sigma_{\theta}^2g(\mathbf{a}))\right)\;,
\end{equation}
where $g(\mathbf{a})=\mathbf{a}^H\mathbf{H}^H\mathbf{C}_w^{-1}\mathbf{H}\mathbf{a}$, and in the above derivation we have used the following equality
\begin{align}
\mathrm{ln}(\gamma) &+ \mathrm{ln}\mathrm{det}(\mathbf{C}_s+\mathbf{C}_w)-\mathrm{ln}\mathrm{det}(\mathbf{C}_w)\nonumber\\
=&\;\mathrm{ln}(\gamma)+\mathrm{ln}\mathrm{det}(\mathbf{C}_s\mathbf{C}_w^{-1}+\mathbf{I}_M)\nonumber\\
\overset{(a)}{=}&\;\mathrm{ln}(\gamma)+\mathrm{ln}\left(1+\lambda_{\max}(\mathbf{C}_s\mathbf{C}_w^{-1})\right)\nonumber\\
=&\;\mathrm{ln}\left(\gamma(1+\sigma_{\theta}^2g(\mathbf{a}))\right)\;,
\end{align}
where $(a)$ is due to the fact that $\mathbf{C}_s\mathbf{C}_w^{-1}$ is a rank-one matrix and $\lambda_{\max}(\cdot)$ is the largest eigenvalue of its matrix argument. Using the matrix inversion lemma, the left hand side of (\ref{eq:loglr}) is calculated as
\begin{equation}\label{eq:matrixinv}
\mathbf{C}_w^{-1}-(\mathbf{C}_s+\mathbf{C}_w)^{-1}=\frac{\sigma_{\theta}^2}{1+\sigma_{\theta}^2g(\mathbf{a})}\mathbf{C}_w^{-1}\mathbf{H}\mathbf{a}\mathbf{a}^H\mathbf{H}^H\mathbf{C}_w^{-1}\;,
\end{equation}
and substituting (\ref{eq:matrixinv}) into (\ref{eq:loglr}) will produce the desired result.

\section{Proof of Theorem 2}\label{app:theorem2}
Beginning with the low transmit power case, assume the following suboptimal choice for the transmission gains: $|\bar{a}_{i}|=\sqrt{\frac{\sigma_n^2 d_i^{\alpha}}{2\sigma_{v,i}^2M}}$, which results in
\begin{equation}
\label{eq:psum}
P=\sum_{i=1}^N|\bar{a}_i|^2=\frac{1}{2M}\sum_{i=1}^{N}\frac{\sigma_n^2 d_i^{\alpha}}{\sigma_{v,i}^2} = O(1/M) \; ,
\end{equation}
and hence $P\rightarrow 0$ as $M\rightarrow \infty$.  Substituting $|\bar{a}_{i}|$ into~(\ref{eq:approx2}), we have
\begin{equation}
g(\mathbf{\bar{a}})=\frac{1}{3}\sum_{i=1}^N\frac{1}{\sigma_{v,i}^2}\;,
\end{equation}
where $\mathbf{\bar{a}}=[\bar{a}_1\cdots\bar{a}_N]^T$.  The value for $g(\mathbf{\bar{a}})$ can serve as a lower bound for $g(\mathbf{a})$ when evaluated at the optimal solution $\mathbf{a}^*$ obtained using (\ref{eq:optsol}) and using $P$ in (\ref{eq:psum}) as the power constraint:
\begin{equation}\label{eq:glb}
g(\mathbf{a}^{*})\ge \frac{1}{3}\sum_{i=1}^N\frac{1}{\sigma_{v,i}^2}\;.
\end{equation}
Substituting (\ref{eq:glb}) into (\ref{eq:optpd}), we have the lower bound for the multi-antenna FC:
\begin{equation}\label{eq:lbm}
P_{D}\ge \epsilon^{\frac{1}{1+\frac{\sigma_{\theta}^2}{3}\sum_{i=1}^N\frac{1}{\sigma_{v,i}^2}}} > \epsilon \; .
\end{equation}
For the single-antenna FC, based on (\ref{eq:optrs}) we have the following upper bound since $\frac{P}{\sigma^2_n}\mathbf{I}_N\succeq\mathbf{R}^{-1}$:
\begin{equation}\label{eq:upls}
\frac{\sigma_s^2}{\sigma_w^2}{\le}\frac{\sigma_{\theta}^2P}{\sigma_n^2}\mathbf{h}^H\mathbf{h}\;.
\end{equation}
Using (\ref{eq:upls}) and (\ref{eq:psum}) together with~(\ref{pfad}) and~(\ref{pfas}), it is easy to show that
\begin{equation}\label{eq:uplsp}
P_{D}^s\le \epsilon^{\frac{1}{1+\zeta}},
\end{equation}
where $\zeta=\frac{1}{2M}\sum_{i=1}^N\frac{\sigma_{\theta}^2d_i^{\alpha}}{\sigma_{v,i}^2}\mathbf{h}^H\mathbf{h}$. According to the Rayleigh channel model, $\mathbf{h}^H\mathbf{h}$ is the sum of weighted chi-squared random variables, and for an arbitrary positive number $\tau$ we have
\begin{align}
\lim_{M\to\infty}\mathrm{Pr}\left(\zeta>\!\tau\right)\le\lim_{M\to\infty}\mathrm{Pr}\!\left(\!\!\frac{\sigma_{\theta}^2N}{4M}\frac{\max_i\frac{d_i^{\alpha}}{\sigma_{v,i}^2}}{\min_i d_i^{\alpha}}\chi_{(2N)}^2\!>\!\tau\!\!\right)=0\;,
\end{align}
where $\chi_{(2N)}^2$ denotes a chi-square variable with $2N$ degrees of freedom. Thus, $\zeta$ converges to $0$ in probability and hence $P_{D}^s$ converges to $\epsilon$ in probability.

From (\ref{eq:approx2}), it is clear that for very large $M$, $g(\mathbf{a})$ is upper bounded by
\begin{equation}\label{eq:ub}
g(\mathbf{a})\le\sum_{i=1}^N\frac{1}{\sigma_{v,i}^2}\;.
\end{equation}
Note that the lower bound in (\ref{eq:glb}) is one third the upper bound in (\ref{eq:ub}). When $P\to\infty$ and hence $|a_i|$ is large, the upper bound in~(\ref{eq:ub}) can be asymptotically achieved even with an equal power allocation $|a_i|=\sqrt{P/N}$. Also, we see that to maximize the upper bound for $g(\mathbf{a})$ in this case, all the sensors should transmit. Plugging~(\ref{eq:ub}) into~(\ref{eq:optpd}), we have the following upper bound for $P_D$:
\begin{equation}\label{eq:ubm}
P_{D}\le \epsilon^{\frac{1}{1+\sigma_{\theta}^2\sum_{i=1}^N\frac{1}{\sigma_{v,i}^2}}}\;.
\end{equation}
For the single-antenna FC, according to~(\ref{eq:optrs}), we have the following bound as $P \rightarrow \infty$ since $(\mathbf{F}\mathbf{V}\mathbf{F}^H)^{-1}\succeq\mathbf{R}^{-1}$:
\begin{equation}\label{eq:ubrs}
\frac{\sigma_s^2}{\sigma_w^2}\le\sigma_{\theta}^2\sum_{i=1}^N\frac{1}{\sigma_{v,i}^2}\; .
\end{equation}
Using~(\ref{eq:ubrs}) together with~(\ref{pfad}) and~(\ref{pfas}) yields
\begin{equation}\label{eq:ubs}
P_{D}^s\le \epsilon^{\frac{1}{1+\sigma_{\theta}^2\sum_{i=1}^N\frac{1}{\sigma_{v,i}^2}}}\;.
\end{equation}
Note that for both~(\ref{eq:ubm}) and (\ref{eq:ubs}), the inequality is asymptotically achieved as $P\to\infty$, which completes the proof.

\section{Proof of Theorem 3}\label{app:theorem3}
\renewcommand{\theequation}{\thesection.\arabic{equation}}

Using the definition in (\ref{eq:deflec}),
\begin{align}\label{eq:deflec1}
\lim_{M\to\infty}D\left(T\right)=\frac{(\mu_{e,1}-\mu_{e,0})^2}{\sigma_{e,0}^2}=\frac{\sigma_{\theta}^4\left(\sum_{i=1}^N\frac{|a_i|^2}{d_i^{\alpha}}\right)^2}{\sum_{i=1}^N\left(\frac{\sigma_{v,i}^2|a_i|^2}{d_i^{\alpha}}+\frac{\sigma_n^2}{M}\!\right)^2+\frac{M-N}{M^2}\sigma_n^4}\;,
\end{align}
where the parameters $\mu_{e,1}$, $\mu_{e,0}$ and $\sigma_{e,0}^2$ are defined and calculated below.  For $\mu_{e,1}$,
\begin{align}
\mu_{e,1}&=\lim_{M\to\infty}\mathbb{E}\left\{\left.\frac{1}{M}\mathbf{y}^H\mathbf{y}\right|\mathcal{H}_1\right\}\nonumber\\
&=\lim_{M\to\infty}\frac{1}{M}\mathbb{E}\left\{\tilde{\mathbf{y}}^H(\mathbf{C}_w+\mathbf{C}_s)\tilde{\mathbf{y}}\right\}\nonumber\\
&=\lim_{M\to\infty}\frac{1}{M}\mathrm{tr}\left(\mathbf{C}_w+\mathbf{C}_s\right)\nonumber\\
&=\lim_{M\to\infty}\frac{1}{M}\mathrm{tr}\left(\sigma_{\theta}^2\mathbf{H}^H\mathbf{H}\mathbf{a}\mathbf{a}^H\!+\mathbf{H}^{H}\mathbf{H}\mathbf{D}\mathbf{V}\mathbf{D}^{H}\right)+\sigma_n^2\nonumber\\
&\overset{(j)}{=}\sum_{i=1}^N\frac{\left(\sigma_{\theta}^2+\sigma_{v,i}^2\right)|a_i|^2}{d_i^{\alpha}}+\sigma_n^2\;,
\end{align}
where $\tilde{\mathbf{y}}$ has distribution $\mathcal{CN}\left(\mathbf{0},\mathbf{I}_M\right)$ and in $(j)$ we used (\ref{eq:approx}). Similarly, we have
\begin{align}
\mu_{e,0}&=\lim_{M\to\infty}\mathbb{E}\left\{\left.\frac{1}{M}\mathbf{y}^H\mathbf{y}\right|\mathcal{H}_0\right\}\nonumber\\
&=\sum_{i=1}^N\frac{\sigma_{v,i}^2|a_i|^2}{d_i^{\alpha}}+\sigma_n^2\;,\\
\sigma_{e,0}^2&=\lim_{M\to\infty}\mathrm{Var}\left\{\left.\frac{1}{M}\mathbf{y}^H\mathbf{y}\right|\mathcal{H}_0\right\}\nonumber\\
&=\lim_{M\to\infty}\frac{1}{M^2}\mathrm{Var}\{\tilde{\mathbf{y}}^H\mathbf{C}_w\tilde{\mathbf{y}}\}\nonumber\\
&\overset{(h)}{=}\lim_{M\to\infty}\frac{1}{M^2}\mathrm{tr}(\mathbf{C}_w^2)\nonumber\\
&=\lim_{M\to\infty}\sum_{i=1}^N\left(\frac{\sigma_{v,i}^2|a_i|^2}{d_i^{\alpha}}+\frac{\sigma_n^2}{M}\right)^2\!+\frac{(M-N)}{M^2}\sigma_n^4\;,
\end{align}
where in $(h)$ we used the following lemma proved in Appendix~\ref{app:lemma2}:
\begin{lemma}\label{eq:lemma2}
Given a complex Gaussian random vector $\mathbf{z}\in \mathbb{C}^{M\times 1}$ with distribution $\mathcal{CN}(\mathbf{0},\mathbf{I}_M)$, and a Hermitian matrix $\mathbf{A}\in\mathbb{C}^{M\times M}$,  the variable
$\mathbf{z}^H\mathbf{A}\mathbf{z}$ has a variance $\mathrm{Var}\{\mathbf{z}^H\mathbf{A}\mathbf{z}\}=\mathrm{tr}(\mathbf{A}^2)$\;.
\end{lemma}

Introducing new variables $x_i=|a_i|^2$, (\ref{eq:deflec1}) is equivalent to
\begin{align}\label{eq:deflec2}
\lim_{M\to\infty}D\left(T\right)&=\frac{\sigma_{\theta}^4\left(\sum_{i=1}^N\frac{x_i}{d_i^{\alpha}}\right)^2}{\sum_{i=1}^N\left(\frac{\sigma_{v,i}^2x_i}{d_i^{\alpha}}+\frac{\sigma_n^2}{M}\right)^2+\frac{M-N}{M^2}\sigma_n^4}\nonumber\\
&=\frac{\sigma_{\theta}^4\mathbf{x}^T\mathbf{d}\mathbf{d}^T\mathbf{x}}{\mathbf{x}^T\mathbf{B}\mathbf{x}+\frac{2\sigma_n^2}{M}\mathbf{b}^T\mathbf{x}+\frac{\sigma_n^4}{M}}\;,
\end{align}
where the variables $\mathbf{x, d, B, b}$ are defined in~(\ref{eq:d1})-(\ref{eq:d2}). Substituting $x_i=\frac{P_i}{\sqrt{M}}$ into (\ref{eq:deflec2}), we obtain
\begin{align}
\lim_{M\to\infty}D\left(T\right)&=\lim_{M\to\infty}\frac{\sigma_{\theta}^4\mathbf{p}^T\mathbf{d}\mathbf{d}^T\mathbf{p}}{\mathbf{p}^T\mathbf{B}\mathbf{p}+\frac{2\sigma_n^2}{\sqrt{M}}\mathbf{b}^T\mathbf{p}+\sigma_n^4}\nonumber\\
&=\frac{\sigma_{\theta}^4\mathbf{p}^T\mathbf{d}\mathbf{d}^T\mathbf{p}}{\mathbf{p}^T\mathbf{B}\mathbf{p}+\sigma_n^4}\;,\label{eq:sysdefl}
\end{align}
where $\mathbf{p}=[P_1\cdots P_N]$, and we see that $D(T)$ is asymptotically independent of $M$.  We also observe from~(\ref{eq:deflec2}) that an asymptotically non-zero deflection requires that $|a_i|^2$ not decrease faster than $\frac{1}{\sqrt{M}}$.

\section{Proof of Lemma 2}\label{app:lemma2}
We first rewrite $\mathbf{z}^H\mathbf{A}\mathbf{z}$ as
\begin{equation}
\mathbf{z}^H\mathbf{A}\mathbf{z}=\sum_{i=1}^M \frac{\lambda_i(\mathbf{A})}{2} \chi_{i}^2(2)\;,
\end{equation}
where $\lambda_i(\mathbf{A})$ are the eigenvalues of $\mathbf{A}$ and $\chi_{i}^2(2)$ are independent chi-squared variables with 2 degrees of freedom, which can be expressed as
\begin{equation}
\chi_{i}^2(2)=z_{i,1}^2+z_{i,2}^2\;,
\end{equation}
where the independent variables $z_{i,1}$ and $z_{i,2}$ have normal distribution $\mathcal{N}(0,1)$. Since $\mathbf{z}^H\mathbf{A}\mathbf{z}$ can be viewed as the sum of $M$ independent variables, the variance of $\mathbf{z}^H\mathbf{A}\mathbf{z}$ is calculated as
\begin{align}
\mathrm{Var}\{\mathbf{z}^H\mathbf{A}\mathbf{z}\}&=\sum_{i=1}^M \frac{\lambda_i^2(\mathbf{A})}{4} \mathrm{Var}\{\chi_{i}^2(2)\}\nonumber\\
&=\sum_{i=1}^M \frac{\lambda_i^2(\mathbf{A})}{4}\left(\mathrm{Var}\{z_{i,1}^2\}+\mathrm{Var}\{z_{i,2}^2\}\right)\nonumber\\
&\overset{(u)}{=}\sum_{i=1}^M \lambda_i^2(\mathbf{A})\nonumber\\
&\overset{(t)}{=}\mathrm{tr}(\mathbf{A}^2)\;,
\end{align}
where $(u)$ follows from
\begin{equation}
\mathrm{Var}\{z_{i,k}^2\}=\mathbb{E}\{z_{i,k}^4\}-\left(\mathbb{E}\{z_{i,k}^2\}\right)^2=2\;,
\end{equation}
and $(t)$ is due to the fact that $\lambda_i^2(\mathbf{A})$ are the eigenvalues of the matrix $\mathbf{A}^2$.



\bibliography{reference}
\bibliographystyle{IEEEbib}

\newpage




\end{document}